\documentclass[epj]{svjour}

\usepackage{graphicx}

\begin{document}

\newcommand{\extr}{\mathop{\rm extr}}
\newcommand{\J}{$J$}
\newcommand{\Jo}{$J_{0}$}
\newcommand{\Tf}{$T_{f}$}
\newcommand{\Jk}{$J_{K}$}
\newcommand{\Jkc}{$J_{K}^{c}$}
\newcommand{\TsJ}{$T/J$}
\newcommand{\JosJ}{$J_{0}/J$}
\newcommand{\JksJ}{$J_{K}/J$}
\newcommand{\cao}{\c c\~ao}

\renewcommand{\thefootnote}{\alph{footnote}}

\title{Spin glass and Ferromagnetism in Kondo lattice compounds} 

\author{S.\ G.\ Magalh\~aes\inst{1,\mbox{\scriptsize a}} \and 
        A.\ A.\ Schmidt\inst{2,\mbox{\scriptsize b}} \and 
	Alba Theumann\inst{3,\mbox{\scriptsize c}} \and 
	B.\ Coqblin\inst{4,\mbox{\scriptsize d}}}

\institute{Departamento de F\'\i sica -- UFSM, 97105-900 Santa Maria, RS, Brazil \and
           Departamento de Matem\'atica -- UFSM, 97105-900 Santa Maria, RS, Brazil \and
	   Instituto de F\'\i sica -- UFRGS, 91501-970 Porto Alegre, RS, Brazil \and
	   Laboratoire de Physique des Solides, Universit\'e Paris-Sud, 91405 Orsay, France}

\date{Received: ??? / Revised version: ???}

\abstract{
The Kondo lattice model has been analyzed in the presence of a 
random inter-site interaction among localized spins with non 
zero mean \Jo\ and standard deviation \J. Following the same 
framework previously introduced  by us, the problem is 
formulated in the path integral formalism where the spin operators are 
expressed as bilinear combinations of Grassmann fields. The static 
approximation and the replica symmetry ansatz have allowed us to 
solve the problem at a mean field level. The resulting phase 
diagram displays several phase transitions among a 
ferromagnetically ordered region,a spin glass one, a mixed phase and a
Kondo state depending on \Jo, \J\ and its relation with the 
Kondo interaction coupling \Jk. These results could be used to address 
part of the experimental data for the $CeNi_{1-x}Cu_x$ compound, 
when $x \leq 0.8$.
\PACS{
      {64.60.Cn}{Order-disorder transformations; statistical mechanics of model systems} \and
      {75.10.Nr}{Spin-glass and other random models } \and
      {75.30.Mb}{Valence fluctuation, Kondo lattice, and heavy-fermion phenomena} 
     } 
} 

\maketitle

\footnotetext[1]{\email{ggarcia@ccne.ufsm.br}}
\footnotetext[2]{\email{alex@lana.ccne.ufsm.br}}
\footnotetext[3]{\email{albath@if.ufrgs.br}}
\footnotetext[4]{\email{coqblin@lps.u-psud.fr}}

\section{Introduction} \label{intro}

The magnetism in strongly correlated f-electron systems has become a source 
of great interest due to the physics involved \cite{r1} like, for instance, 
quantum phase transitions and Non-Fermi liquid behavior \cite{r2}. The 
anti-ferromagnetic s-f exchange coupling of conduction electrons to 
localized spins can be responsible for the competition between the Kondo 
effect,that reduces the localized magnetic moments, and the RKKY 
interaction among magnetic impurities which, in turn, may give rise to  
magnetic long range order.  

Recently, an experimental magnetic phase diagram of the Kondo 
$CeNi_{1-x}Cu_{x}$ compound has been proposed \cite{r3} showing the 
existence of a spin glass like state. In the $CeCu$ limit, the negative 
magnetic interaction is dominant enough to produce an anti-ferromagnetic 
long range order with no indications of the Kondo effect. When $Cu$ is 
substituted by $Ni$, there is a phase transition around 
$x=0.8$ from the antiferromagnetic (AF) to a ferromagnetic (FM) ordering,
which finally disappears at roughly $x=0.2$; the Curie temperature is 
roughly equal to 1K and is slowly decreasing down to $x=0.4$ and then
disappears at $x=0.2$. Above the ferromagnetic phase, a spin-glass (SG)
phase was identified by magnetic susceptibility measurements and the SG 
transition temperature increases from 2 to 6K for $x$ varying from
0.7 to 0.2. For $x<0.2$ a Kondo behaviour has been proposed, and finally 
$CeNi$ is an intermediate valence compound. Thus, at very low temperatures,
the phase sequence FM-SG-Kondo has been observed with 
decreasing $x$ and in the range $0.7-0.2$ for $x$, the sequence FM-SG is 
obtained with increasing temperature. It is quite rare to observe a 
ferromagnetic phase in Cerium Kondo compounds, while antiferromagnetic 
phases are often observed and for example the sequence of SG-AF- Kondo 
transitions is obtained with increasing $x$ in 
$Ce_{2}Au_{1-x}Co_{x}Si_{3}$ alloys \cite{r4}. 

Quite recently, a model has been introduced \cite{r5} to study the 
interplay between spin glass ordering and a Kondo state. This model 
is based on the previously introduced Kondo lattice model \cite{r6} with 
an intrasite s-f exchange interaction and an intersite long range random 
interaction of zero mean that couples the localized spins. The use of the 
static approximation and the replica symmetry ansatz has made possible 
to solve the problem at a mean field level. This fermionic problem is 
formulated by representing the spin operators as a bilinear combination 
of Grassmann fields and the partition function is found through the 
functional integral formalism \cite{r7,r8,r9,r10}. The results are shown 
in a phase diagram of \TsJ\ {\it versus} \JksJ\ where $T$ is the 
temperature, \Jk\ is the intrasite Kondo exchange interaction and $J$
is the standard deviation of the random inter-site interaction. For high 
temperatures and small values of \Jk, a paramagnetic phase is found. 
In this situation, if the temperature is decreased a second order 
phase transition to a spin glass phase appears at \Tf. The model shows 
a transition line \Jkc$(T)$ separating the paramagnetic and the spin 
glass phases from the Kondo phase.

In the present work, the model mentioned in the previous paragraph has 
been extended in order to include the proper elements that produce also a 
ferromagnetic ordering by taking the mean random interaction \Jo\ to be 
different from zero. Therefore, the magnetization can be introduced in 
addition to the other order parameters and solved coupled to them.  

From this procedure a quite non-trivial phase diagram is 
obtained which contains ferromagnetism, a mixed phase \cite{r12,r13} 
(ferromagnetism and spin glass), a spin glass phase and a Kondo state.
For instance, one of the achievements of the present work is the finding 
of a mixed phase whose existence should not be discarded in the magnetic 
measurement of $CeNi_{1-x}Cu_{x}$, as mentioned in Ref.\ \cite{r3}

This paper is organized as follows. In section II we present the model 
and its development in order to obtain the free energy  and the saddle point
coupled equations for the order parameters. The phase diagram of the temperature 
\TsJ\ {\it versus} \JksJ\ is shown for several values of \Jo. 
The Almeida-Thouless line is also calculated. Discussions and concluding 
remarks are presented in the last section.

\section{The model and results} \label{model}

The model considered in this work was introduced before in 
Ref.\ \cite{r5} to study  spin glass ordering in a Kondo lattice 
compound so the Hamiltonian is

\begin{eqnarray} 
\displaystyle
{\cal H}-\mu_{c}N_{c}-\mu_{f}N_{f}= \sum_{k,\sigma}\epsilon_{k}n_{k\sigma}+ 
 \epsilon_{0}\sum_{i,\sigma }n_{i\sigma }^{f} + \nonumber\\
 J_{K} \sum_{i}[S_{fi}^{+}s_{i}^{-}+S_{fi}^{-}s_{i}^{+}] - 
\sum_{i,j}J_{ij}S_{fi}^{z}S_{fj}^{z}
\label{e1}
\end{eqnarray}

\noindent where $J_{K} > 0$ and the sum runs over N lattice sites. In the 
present case the random intersite  interaction $J_{ij}$  in the Hamiltonian 
is infinite ranged with a Gaussian distribution where $<J_{ij}> = 2J_0/N$ 
and $<J_{ij}^2> = 8J^2/N$. This particular scaling compensates the factors 
$1/2$ that originate in the definition of the operators $S^z$ in equation (\ref{e2})
and also in changing from sum over sites to sum over bonds.
 

The spin variables $S^{(+-)}_{fi}$ $(s^{(+-)}_{ci})$, $S^{z}_{fi}$ are 
bilinear combinations of the creation and destruction operators \cite{r5} 
for localized  (conduction) fermions $f_{i\sigma}^{\dagger}$, $f_{i\sigma}$ 
($d_{i\sigma}^{\dagger}$,$d_{i\sigma}$)  with the spin projection 
$\sigma=\uparrow$ or $\downarrow$:

\begin{eqnarray}
{S_{fi}^{+}}=f_{i\uparrow}^{\dagger}f_{i\downarrow}\hspace{0.8cm};
\hspace{0.8cm}s_{ci}^{+}=d_{i\uparrow}^{\dagger}
d_{i\downarrow}\nonumber\\
{S_{fi}^{-}}=f_{i\downarrow}^{\dagger}f_{i\uparrow}\hspace{0.8cm};
\hspace{0.8cm}s_{ci}^{-}=d_{i\downarrow}^{\dagger}d_{i\uparrow}\nonumber\\
{S_{fi}^{z}}=\frac{1}{2}[f_{i\uparrow}^{\dagger}f_{i\uparrow}-
f_{i\downarrow}^{\dagger}f_{i\downarrow}]
\label{e2}
\end{eqnarray}

The $\mu_f$ ($\mu_c$) are the chemical potential for the localized  
(conduction) band. The energy $\epsilon_0$ is referred to $\mu_f$ 
while $\epsilon_k$ is referred to $\mu_c$. 

The partition function is expressed in terms of functional integrals using 
anticommuting Grassmann variables $\varphi_{i\sigma}(\tau)$ and 
$\psi_{i\sigma }(\tau)$ associated with the conduction and the localized 
electrons respectively. Therefore,

\begin{eqnarray}
Z  = \int D(\psi^{\ast}\psi) D(\varphi^{\ast}\varphi)
\exp\left\{ \int_0^{\beta}\!d\tau
\left[ L_0(\psi^{\ast}\!,\psi)\right.\right.+ \nonumber \\ 
\left. \phantom{\int_0^{\beta}} \left.  L_0(\varphi^{\ast}\!,\varphi) +
      L_{SG} + L_K \right] \right\}
\label{e3}
\end{eqnarray}
where
\begin{eqnarray}
L_0(\psi^{\ast}\!,\psi) & = &\sum_{ij\sigma} \psi_{i\sigma}^{\ast}(\tau)
\left[ \frac{\partial}{\partial\tau} - \varepsilon_{0}\right]
\delta_{ij}\psi_{j\sigma}(\tau),~ \nonumber \\
L_0(\varphi^{\ast}\!,\varphi) & = & \sum_{ij\sigma} \varphi_{i\sigma}^{\ast}(\tau)
\left[ \frac{\partial}{\partial\tau}\delta_{ij} - t_{ij} 
\right]\,\varphi_{j\sigma}(\tau), \nonumber \\
L_{SG} & = &\sum_{ij} J_{ij} S_{fi}^{z}(\tau) 
S_{fj}^{z}(\tau), \nonumber \\ 
L_K & = &\frac{J_K}{N} \sum_{i\sigma} 
\left[ \varphi_{i-\sigma}^{\ast}(\tau)
\psi_{i-\sigma}(\tau)\right] \times \nonumber \\
& & \phantom{\frac{J_K}{N}} \sum_{j\sigma}
\left[ \psi_{j\sigma}^{\ast}(\tau)
\varphi_{j\sigma}(\tau)\right].
\label{e4}
\end{eqnarray}

In the static approximation \cite{r7,r8,r9,r10}, it is possible to solve the 
problem in a mean field theory where the Kondo state is described by the 
complex order parameters \cite{r5,r6}:

\begin{eqnarray}
\lambda_{\sigma}^{\ast}=\frac{1}{N}\displaystyle \sum_{i,\omega} 
\langle \psi_{i\sigma }^{\ast}(\omega) \varphi_{i\sigma}(\omega) 
\rangle\phantom\nonumber\\
\lambda_{\sigma}=\frac{1}{N}\displaystyle\sum_{i,\omega} 
\langle \varphi_{i\sigma}^{\ast}(\omega)\psi_{i\sigma}(\omega) \rangle
\label{e5}
\end{eqnarray}

Following the treatment for the Kondo part in the partition function as 
introduced in Ref.\ \cite{r5}, where it was assumed that 
$\lambda_{\sigma }^{\ast}\approx\lambda^{\ast}$ 
$(\lambda_{\sigma }\approx\lambda)$, we show in the Appendix that
first the conduction electron degrees of freedom may be integrated
out to give

\begin{eqnarray}
\frac{Z}{Z^{0}_{d}}\,=\,e\,^{[-2N\beta J_{K}\lambda^{\ast}\lambda]}\,Z_{eff}
\label{e6}
\end{eqnarray}
where
\begin{eqnarray}
Z_{eff}\!=\!\int\!D(\psi^{\ast}\psi)\,\exp\left\{\sum_{\omega\sigma}
\sum_{i,j}g_{ij}^{-1}(\omega)\psi^{\ast}_{i\sigma}(\omega)
\psi_{j\sigma}(\omega) \right.\nonumber \\
\left.\phantom{\sum_{ij\omega\sigma}}+A_{SG}\right\},~
\label{e7}
\end{eqnarray}
$Z^{0}_{d}$ is the partition function of the free conduction electrons,
\begin{eqnarray}
g^{-1}_{ij}(\omega)=(i\omega-\beta\epsilon_0)\delta_{ij}
-\beta^{2}J_{K}^{2}\lambda^{\ast}\lambda\gamma_{ij}(\omega),
\label{e8}
\end{eqnarray}
while $\gamma^{-1}_{ij}=i\omega\delta_{ij}-\beta t_{ij}$ is the inverse 
d-electron Green's function and $\beta=1/T$ is the inverse temperature.

The free energy is given by the replica method 
\begin{eqnarray}
\beta F= 2\beta J_{K}\lambda^{\ast}\lambda -
\lim_{n\rightarrow 0}\frac{1}{Nn}
(\langle \langle Z^{n}_{eff}(J_{ij}) \rangle \rangle_{ca}-1)
\label{e9}
\end{eqnarray}
and the averaged replicated partition function can be linearized by 
means of the usual Hubbard-Stratonovich \linebreak transformation. Therefore,
\begin{eqnarray}
\langle\langle Z^{n}_{eff}(J_{ij}) \rangle\rangle_{ca} =
\int\Pi_{\alpha \beta}dq_{\alpha\beta}\int\Pi_{\alpha}dm_{\alpha}
\exp\left\{-N\times\phantom{\sum_{\alpha\beta}}\right.\hspace*{-10mm}\nonumber\\
\left\{\left.\frac{\beta^{2}J^{2}}{2}\sum_{\alpha\beta}q_{\alpha\beta}^{2} +
\frac{\beta J_{0}}{2}\sum_{\alpha}m^{2}_{\alpha}\right\}\right\} 
\Lambda({q_{\alpha\beta},m_{\alpha}})~~
\label{e10}
\end{eqnarray}
with $\alpha = 0,1,..n $ being the replica index and 
\begin{eqnarray}
\Lambda({q_{\alpha\beta},m_{\alpha}})=\int D(\psi^{\ast}_{\alpha}\psi_{\alpha})\,
\exp\left\{ \sum_{i,j \sigma,\omega} g^{-1}_{ij}(\omega)\ \times \right.\nonumber\\
\sum_{\alpha} \psi^{\ast}_{i \sigma \alpha}(\omega)\psi_{j 
\sigma\alpha}(\omega)+
\beta J_{0}\sum_{i\alpha}2S^{\alpha}_{i}m_{\alpha}+ \nonumber\\
\left.\beta^{2}J^{2}\sum_{ij\alpha\beta}4S^{\alpha}_{i}S^{\beta}_{j}q_{\alpha
\beta}\right\}. 
\label{e11}
\end{eqnarray}
A more detailed derivation of equations (\ref{e10}) and (\ref{e11}) is 
given in the Appendix.

This problem is analysed within the replica symmetric ansatz 
where $q_{\alpha\neq\beta}=q$ is the spin glass order parameter, 
$m_{\alpha}=m$ is the magnetization and $q_{\alpha\alpha}=
q+\overline{\chi}$, ($\overline{\chi}=\frac{\chi}{\beta}$) with $\chi$
being the static susceptibilty. The sum over replica indices also gives 
quadratic terms which can be linearized again by introducing new auxiliary 
fields  in equation (\ref{e10}):

\begin{eqnarray}
\Lambda({q_{\alpha\beta},m_{\alpha}})=\int Dz_j \int D(\psi^{\ast}\psi)\, 
\times \nonumber\\
\exp\left\{ \sum_{i,j \sigma,\omega} g^{-1}_{ij}
\sum_{\alpha} \psi^{\ast}_{i\sigma\alpha}(\omega_n)\, 
\psi_{j\sigma\alpha}(\omega_n)\,+ \right.\nonumber\\ 
\left. \beta\,J_{0}\,m \sum_{\alpha} 2S_{i}^{\alpha} + 
\beta\,J\,\sqrt{2q} \sum_{i\phantom{j}}\,z_i\,2S_i^{\alpha} \right\} 
\times \nonumber\\
\int D \xi_j^{\alpha}\,\exp\left\{ - \sum_{\alpha j}\,(\xi_j^{\alpha})^2
+ \beta J \sqrt{2\overline{\chi}}\,\xi_j^{\alpha}\,2S_j^{\alpha}\right\}
\label{e12}
\end{eqnarray}
where $Dx=\frac{1}{\sqrt{2\pi}}e^{-x^{2}/2}dx$ and 
\[ S^{\alpha}_i = \frac{1}{2} \sum_{\omega_n\alpha\sigma=\pm} \sigma 
\psi^{\ast}_{\alpha\sigma}(\omega_n) \psi_{\alpha\sigma}(\omega_n)\ . \]
 
The functional integral in equation (\ref{e12}) can be performed and the saddle 
point solution for the free energy is given by

\begin{eqnarray}
\beta F\,\,=\,\,
2\beta J_{K}\lambda^{2}\,+\,\frac{1}{2}\beta^{2}J^{2}
\left(\overline{\chi}^2+2q\overline{\chi}\right)+\nonumber\\
\frac{\beta J_{0}}{2}m^{2}-\lim_{n\rightarrow 0}\frac{1}{Nn}\left\{
\int \Pi_{i} Dz_{i}\ \times \right.\nonumber\\
\int\Pi_{\alpha,i} D\xi_{i}^{\alpha}
\left.\exp\left[\sum_{\omega\sigma}\ln\left[\det G_{ij\sigma}^{-1}(\omega)
\right]\right]-1\right\}.
\label{e13}
\end{eqnarray}
where in the previous equation we introduced the inverse Green's function
\begin{eqnarray}
G^{-1}_{ij\sigma}(\omega)=
g_{ij}^{-1}(\omega)-\delta_{ij}\sigma h(z_{i},\xi_{i}^{\alpha})
\label{e14}
\end{eqnarray}
with an effective field
\begin{eqnarray}
h(z_{i},\xi_{i}^{\alpha})=\beta J_{0}m+
\beta J \sqrt{2q}z_{i}+\beta J \sqrt{2\overline{\chi}}\xi^{\alpha}_{i}.
\label{e15}
\end{eqnarray}

A problem is presented by the calculation of the \linebreak Green's function 
$G_{ij\sigma}(\omega)$ in equation (\ref{e14}) where there is a random Gaussian 
field $h(z_{i},\xi_{i}^{\alpha}) \equiv h_{i\alpha}$ applied at every site 
$i$ of $n$ replicated lattices with $N$ sites. The decoupling used 
at this point is the same as for Ref.\ \cite{r5}, i.e., 
the original Green's function $G_{ij\sigma}(\omega)$ is replaced by 
a Green's function $\Gamma_{\mu\nu\sigma}(\omega)$ where there 
is a uniform field $h_{i\alpha}$ applied in every site $ \mu,\nu$ 
of a ficticious Kondo lattice. Therefore, going to the reciprocal space 
and assuming a constant density of states for the conduction electron 
band $\rho(\epsilon)=\frac{1}{2D}$ for $-D<\epsilon<D$, the sum over 
Matsubara's frequencies $\omega$ can be performed in equation (\ref{e13}) 
and the resulting free energy is 
\begin{eqnarray} 
\beta F\,\,=\,\,2\beta J_{K}\lambda^{2}\,+\,\frac{1}{2}\beta^{2}J^{2}
\left\{\overline{\chi}^{2}+2\overline{\chi}q\right\}\,+
\frac{\beta J_0}{2}m^{2}-~~ \nonumber\\
\int_{-\infty}^{+\infty}Dz\,\ln\left[\int_{-\infty}^{+\infty}D\xi\,\,
e^{E(\xi)}\,\right]~~
\label{e16}
\end{eqnarray}
where
\begin{eqnarray} 
E(\xi) & = &\frac{1}{\beta D}\int_{-\beta D}^{+\beta D}dx
\ln{\{S(\xi,x)\}},\nonumber \\ 
S(\xi,x) & = & \cosh(h_x^{+})+\cosh(\sqrt{\Delta}),\nonumber \\
\Delta & = & (h_x^{-})^2+(\beta J_{k}\lambda)^2 
\mbox{~~and~~} h_x^{\pm} = \frac{h\pm x}{2}. \nonumber
\end{eqnarray}
The saddle point equations for the order parameters $q$, $m$, 
$\overline{\chi}$ and $\lambda$ can be found from equation (\ref{e16}). 

The limit of stability for the order parameters solutions with replica 
symmetry is achieved if the Almeida-Thouless eigenvalue becomes negative:

\begin{eqnarray} 
\lambda_{AT}=1-2\,\beta^{2} J^{2} \int_{-\infty}^{+\infty}
Dz\left\{ \frac{\int_{-\infty}^{+\infty} D\xi\ e^{E(\xi)}\ 
\Omega(T)}{\int_{-\infty}^{+\infty} D\xi\ e^{E(\xi)}}\ -  
\right.\nonumber\\[3mm]
\left.\left[\frac{ \int_{-\infty}^{+\infty} D\xi\ e^{E(\xi)}\ T_1(\xi)}{ 
\int_{-\infty}^{+\infty} D\xi\ e^{E(\xi)}}\right]^2\right\}^2~~
\label{e17} 
\end{eqnarray}     
where
\begin{eqnarray}
\Omega(T) & = & [T_1(\xi)]^2-T_2(\xi)+T_3(\xi)\nonumber \\[3mm] 
T_1(\xi) & = & \frac{1}{2 \beta D}\int_{-\beta D}^{\beta D}dx\, 
\left[\frac{\sinh(h_x^{+})+\frac{\sinh(\sqrt{\Delta})}{\sqrt{\Delta}}
(h_x^{-})}{S(\xi,x)}\right]\ \nonumber \\[3mm] 
T_2(\xi) & = & \frac{1}{4 \beta D}\int_{-\beta D}^{\beta D}dx\,
\left[\frac{ \sinh(h_x^{+})+
\frac{\sinh(\sqrt{\Delta})}{\sqrt{\Delta}}
(h_x^{-})}{ S(\xi,x)}\right]^2\nonumber \\[3mm]
T_3(\xi) & = & \frac{1}{4 \beta D}\int_{-\beta D}^{\beta D}dx\, 
\left[\frac{\cosh(h_x^{+})+\frac{\sinh(\sqrt{\Delta})}
{\sqrt{\Delta}}}{S(\xi,x)} + \right. \nonumber\\
& & \hspace*{10mm}4s\left.\frac{\left[\cosh(\sqrt{\Delta})-
\frac{\sinh(\sqrt{\Delta})}{\sqrt{\Delta}}\right]
\frac{\displaystyle(h_x^{-})^2}{\displaystyle\Delta}}
{S(\xi,x)}\right].\nonumber
\end{eqnarray} 
  
\section{Discussion} \label{discussion}

%
%

\begin{figure}[t] 
\includegraphics[bb=1 1 485 326, scale=0.50]{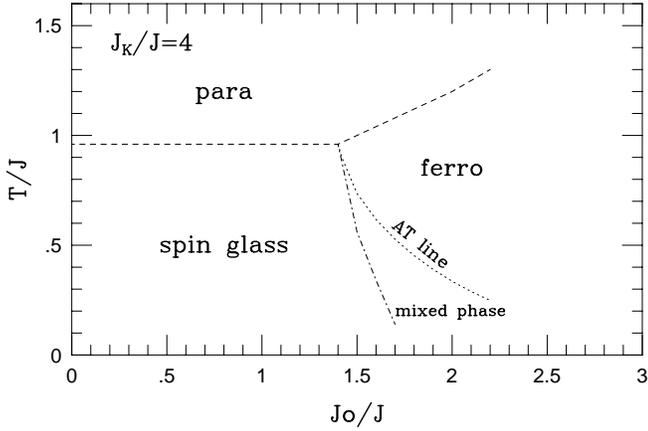}
\caption{Cut in the phase transition space transversal to the \JksJ\ 
axis for \Jk$=2$, $J=0.5$ and $D=10$. The Kondo state is not turned on yet 
and the transitions among the ferromagnetism and the spin glass phases 
depend on the value of \Jo. The dashed line shows the transition from the 
paramagnetic phase to the ferromagnetic and the spin glass phases. The 
dotted line is the Almeida-Thouless (AT) line which, for lower values 
of \Jo, coincides with the paramagnetic -- spin glass transition line 
(horizontal dashed line). The dot-dashed and the AT line delimit the 
mixed phase between the ferromagnetic and spin glass phases.} \label{f1}
\end{figure}

\begin{figure*}[t] 
\includegraphics[bb=1 1 1021 631, scale=0.498]{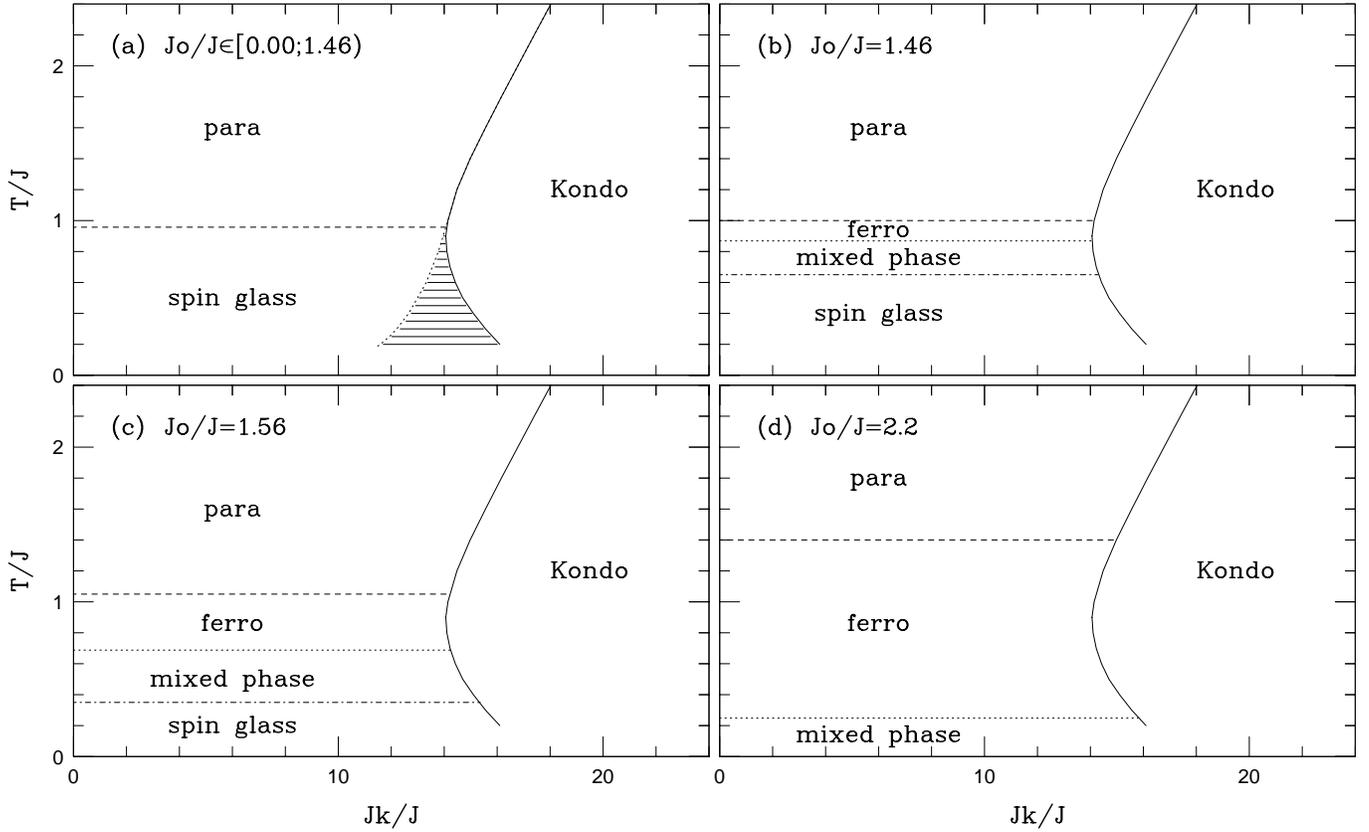}
\caption{Cut in the phase transition space transversal to the \JosJ\ 
axis for several values of \Jo, $J=0.5$ and $D=10$. The solid line shows 
the thermodynamically stable transition from the Kondo phase to the other 
ones. For lower temperatures, a hatched region in panel (a) delineates a 
multiple solution region where the solid line to the right corresponds to 
the thermodynamically stable solution. The horizontal dashed, dotted and 
dot-dashed lines have the same meaning as in Figure \ref{f1}. The AT line 
follows the horizontal dotted line up to the Kondo transition point. Beyond
that, for larger values of \Jk, the AT line follows the transition line 
from the Kondo phase and the mixed and spin glass phases (solid line, lower 
temperatures). One can notice that, as \Jo\  increases, a ferromagnetic and 
a mixed phase start to appear and, for some value of \Jo, the spin glass 
phase finally disappears.} \label{f2}
\end{figure*}

We have studied in this work a Kondo lattice model where the localized 
moments interact through a random inter-site interaction which has an 
average different from zero. The static approximation and replica symmetry 
ansatz lead to a mean field solution for the problem. The resulting coupled 
saddle point equations for the order parameters produce solutions which 
give a Kondo state and magnetic ordering like ferromagnetism, spin glass 
and a mixed phase. In principle, we would be able to build up transition 
surfaces  among those phases in a space \TsJ\ (temperature) {\it versus} 
\JosJ\ (the inter-site interaction average) and \JksJ\ (the Kondo coupling) 
where $J$ (the inter-site interaction standard deviation) is kept constant. 
These parameters \Jk, \Jo, $J$ and the temperature are the set of energy 
scales in the present model. The conduction electrons bandwidth $D$ is kept 
constant.

The result shown in Figure \ref{f1} represents a cut in the cited space 
transversal to the \JksJ\ axis. For this situation and using \JksJ$=4$, 
the Kondo state is still not turned on (it means $\lambda=0$). 
The obtained phase diagram for this fermionic model resembles the 
classical one \cite{r11,r13} (it depends basically on \Jo\ and its relation 
to $J$) except that the numerical values of the transition temperatures 
are smaller. If $J_0<1.46J$, for decreasing temperature, there is a second 
order transition from a paramagnetic to a spin glass phase ($m=0, q\neq0$). 
For that region, the Almeida-Thouless (AT) line coincides with the 
transition line. Howewer, for $J_0\ge1.46J$, for decreasing temperature, 
the model shows a transition from a paramagnetic to a ferromagnetic 
phase ($m\neq0$, $q\neq0$). The AT line is located at higher temperature 
than the calculated replica symmetry line transition $T^*$(\Jo) between 
the ferromagnetic and the spin glass phases. Therefore, this fermionic 
model shows a transition from a ferromagnetic to a replica symmetry 
breaking spin glass phase with a large number of degenerate states but 
still with non-zero spontaneous magnetization which is called a mixed 
phase \cite{r12,r13}.

Figure \ref{f2} shows the cut in the phase space trans\-ver\-sal to 
the \JosJ\ axis. For values of \JosJ\ close to zero (see Figure \ref{f2}a), 
the phase diagram resembles the scenario already found in Ref.\ \cite{r5}, 
that is a paramagnetic phase at high temperatures, a spin glass phase below 
the freezing temperature \Tf\ and a line \Jk=\Jkc(T) separating both phases 
from a Kondo state. For that situation, the AT line is at \Tf\ and 
follows the line \Jk=\Jkc(T). 
           
As the value of \JosJ\ is increased (for small \JksJ) (see Figures \ref{f2}b,
\ref{f2}c and \ref{f2}d), the phase diagram starts to show the presence 
of a ferromagnetic phase which has a transition temperature $T_c$(\Jo) 
increasing with \Jo, the AT line and the calculated replica symmetric 
line $T^*$(\Jo) decreasing with \Jo. Nevertheless the AT line is always 
above $T^*$(\Jo). In that scenario, for decreasing temperature, first a 
transition from paramagnetic to a ferromagnetic phase appears followed by 
a transition from the ferromagnetic to a mixed phase . This behavior is 
reminiscent of that one described for the cut transversal to 
\JksJ\ (Figure \ref{f1}). For some value of \JosJ, the mixed phase finally 
disappears and that region of the phase diagram is totally occupied by the 
ferromagnetic phase. For larger values of \JksJ, the phase diagram goes to 
a Kondo state where the transition line \Jkc(T) does not depend on \Jo.

A remark should be made about the transition line between the spin glass
phase and the Kondo state. At low temperatures, this is a first order 
transition line and so multiple possible solutions for the order parameters can 
be found. Nonetheless, the actual stable solutions can be obtained from 
the minimization of the free energy. In the case of \JosJ$<1.46$ the hatched 
region in Figure \ref{f2}a displays where these multiple solutions occur. By 
computing the free energy we have found the thermodynamically stable solutions.
The solid line to the right of the hatched region in Figure \ref{f2}a corresponds 
to these solutions and the hatched region itself corresponds to metastable 
solutions. Such a carefull analysis can be considered an improvement with 
respect to our previous work \cite{r5} where such a discussion had not been done. 
Nevertheless the previous SG-Kondo state transition line shown there is approximately 
the same as the one presented here. Thus, the hatched regions corresponding to 
the one displayed in Figure \ref{f2}a have not been presented in Figures \ref{f2}b-d.

\section{Conclusions} \label{conclusion}

In this work it has been studied a Kondo lattice model in the 
presence of a random inter-site interaction which \linebreak produces paramagnetism, 
ferromagnetism, a spin glass \linebreak phase, a mixed phase and a region where the 
magnetic moments of the localized electrons are supressed by the screening 
of the conduction ones (Kondo state). The \linebreak model has four energy scales: 
the temperature, \Jo\ (the average inter-site interaction), $J$ 
(the inter-site interaction variance) and \Jk\ (the Kondo coupling). 
As a result one has a three dimensional phase diagram with axes \JosJ, 
\JksJ\ and temperature. Some cuts of this diagram transversal to the 
\JksJ\ and \JosJ, planes are shown in the Figures \ref{f1} and \ref{f2}. 
The position of the transition line separating the Kondo state from the 
magnetic phases is not affected by \Jo. This energy scale is basically 
responsible for locating  several magnetic orderings along the temperature 
range. 

One can try to address the experimental phase diagram found in 
Ref.\ \cite{r3} for the alloys $CeNi_{1-x}Cu_{x}$, but theoretically
if we vary only \Jk\ with $x$, we have found a ferromagnetic phase 
above the spin glass, in disagreement with the experimental result. 
However, the equivalence between their experimental phase diagram and 
ours (see Figures \ref{f1} and \ref{f2}) is not so straightforward since 
the $Ni$ content would have to be associated to both \Jo\ and \Jk. This 
could be an indication that the ergodicity breaking mechanism for the 
formation of magnetic phases like spin glass and ferromagnetism in 
$CeNi_{1-x}Cu_{x}$ is far more complicated than the modelling by a 
random inter-site interaction can address. Although recent investigations 
on the ferromagnetic transverse Ising spin glass suggest also the existence 
of a spin glass transition below the Curie temperature \cite{r14}, it is 
plausible that this be a characteristic of the Sherrington--Kirkpatrick 
model with a high degree of frustration. Less frustrated spin glass 
models \cite{r15} may sustain spin glass order above the Curie temperature 
and they can be more indicated for the study of the $CeNi_{1-x}Cu_{x}$ 
compounds.

To conclude, in Ref.\ \cite{r5} a Kondo lattice model with strong frustation 
(at mean field level) has been solved showing the existence of a SG and a 
Kondo state depending on \JksJ\ (as defined in Section \ref{model}). These 
results could address part of the experimental phase diagram of 
$CeNi_{1-x}Cu_{x}$ \cite{r4}. The purpose of the present work has been to 
examine a wider and more complex region of this experimental phase 
diagram which includes ferromagnetism. Therefore, we have improved our 
previous work by choosing a non-zero average of the random coupling \Jo.
From this approach we have been able to generate a quite non-trivial phase 
diagram with a spin glass phase, ferromagnetism, a Kondo state and a mixed 
phase (spin glass and ferromagnetism). Nevertheless the calculated spin glass
freezing temperature is lower than the Curie temperature in contrast with 
some experimental findings \cite{r3}. However, as pointed out in Ref.\ \cite{r3},
a mixed phase can not be discarded as a possible explanation for the magnetic
measurements. The calculations with the ferromagnetic phase has also shown
an improvement with respect to our previous work \cite{r5} regarding the 
actual location of the SG-Kondo first order transition line. The present 
approach might also explain the frustration in antiferromagnetic Kondo 
systems like $Ce_{2}Au_{1-x}Co_{x}Si_{3}$ alloys. This work is now on progress.


~

\noindent {\small {\bf Acknowledgment} The numerical calculations were  
per\-for\-med at LANA (De\-par\-ta\-men\-to de Ma\-te\-m\'a\-ti\-ca, UFSM) 
and at LSC (Cur\-so de Ci\-\^en\-cia da Com\-pu\-ta\-\c{c}\~ao, UFSM). 
This work was partially supported by the brazilian agencies FA\-PERGS 
(Fun\-da\-\c{c}\~ao\ de Am\-pa\-ro \`a Pes\-qui\-sa do Rio Gran\-de do Sul) 
and CNPq (Con\-se\-lho Na\-cio\-nal de De\-sen\-vol\-vi\-men\-to 
Ci\-en\-t\'\i\-fi\-co e Tec\-no\-l\'o\-gi\-co).}


\section*{Appendix}

We outline here briefly the method used in Ref.\ \cite{r5}.
In order to obtain equations (\ref{e10}) and (\ref{e11}) of Section \ref{model} 
we must first use the static approximation in the Fourier transform of $L_{K}$ 
in equation (\ref{e4}) and introduce the Kondo order parameter by means of the 
identity

\begin{eqnarray} \hspace*{-3mm}
\exp\left\{\int_{0}^{\beta}L_{K}d\tau\right\} & = & \int_{-\infty}^{\infty}
\Pi_{\sigma} d\lambda_{\sigma}^{\dagger}d\lambda_{\sigma}
\Pi_{\sigma} \times \nonumber \\
 & & \delta \left( \phantom{\psi_{j\sigma}^{\dagger}} \hspace*{-5mm}  
\lambda_{\sigma}^{\dagger}N-\right. \sum_{j,w}
\left.\psi_{j\sigma}^{\dagger}(w)\varphi_{j\sigma}(w)\right) \times \nonumber \\
& & \delta \left( \phantom{\varphi_{j\sigma}^{\dagger}} \hspace*{-6mm}
\lambda_{\sigma}N- \right. \sum_{j,w}
\left.\varphi_{j\sigma}^{\dagger}(w)\psi_{j\sigma}(w)\right) \times \nonumber \\
& & \exp\left\{\beta J_{k}N[\lambda_{\uparrow}^{\dagger}\lambda_{\downarrow}+
\lambda_{\downarrow}^{\dagger}\lambda_{\uparrow}]\right\}
\label{a1}
\end{eqnarray}
and using the integral representation for the $\delta$ function we obtain, after
some algebra,

\begin{eqnarray}
Z = \int \Pi_{\sigma}d\lambda^{\dagger}_{\sigma}d\lambda_{\sigma}\, 
\exp\left\{-N\beta J_{K}\sum_{\sigma}
   \lambda^{\dagger}_{\sigma}\lambda_{\sigma}\right\} Z_{stat}~~~~ 
\label{a2}
\end{eqnarray}
where

\begin{eqnarray}
Z_{stat}  = \int D(\psi^{\ast}\psi) 
D(\varphi^{\ast}\varphi) \times \hspace*{34mm} \nonumber \\
\exp \left\{ \int_0^{\beta}\!d\tau\,
\left[ L_0(\psi^{\ast}\!,\psi) + L_0(\varphi^{\ast}\!,\varphi) +
       L_{SG} \right] \right\} \times~~~~ \nonumber \\
\exp\left\{\beta J_{K} \sum_{\sigma} \left[
\lambda_{-\sigma}^{\dagger}
\sum_{j,w}\varphi_{j\sigma}^{\dagger}(w)\psi_{j\sigma}(w) +~~~ 
\right.\right. \nonumber \\
\left.\left.\lambda_{\sigma}
\sum_{j,w}\psi_{j\sigma}^{\dagger}(w)\varphi_{j\sigma}(w)\right]\right\}.~~~~
\label{a3}
\end{eqnarray}

The mean field approximation adopted here is based on two 
assumptions: first, fluctuations in time are ignored (static approximation); 
second, fluctuations in space are also ignored in the definition of
the order parameters. Both assumptions lead us to a quadratic form in the 
Grassmann variables $\varphi$ and $\varphi^{\ast}$ in equation (\ref{a3}) 
and shifting the representation to Matsubara's frequencies we 
can perform the functional integrals to obtain equation (\ref{e6}), where now 
the order parameters $\lambda^{\dagger}_{\sigma}$, $\lambda_{\sigma}$ are taken at
their saddle point value. We also have

\begin{eqnarray}
\int_{0}^{\beta}d\tau L_{SG} = \sum_{ij} J_{ij} S_{fi}^{z} S_{fj}^{z}
\end{eqnarray}
where

\begin{eqnarray}
S_{fi}^{z} = \frac{1}{2}\sum_{\omega} \left[
\psi_{i\uparrow}^{\ast}(\omega)\psi_{i\uparrow}(\omega) -
\psi_{i\downarrow}^{\ast}(\omega)\psi_{i\downarrow}(\omega) \right].
\label{a4}
\end{eqnarray}

Hence, in order to get the configurational average over the random coupling 
\( J_{ij}\), we use a gaussian distribution with average and variance given 
in Section \ref{model}. So

\begin{eqnarray}
\langle Z_{eff}^{n} (J_{ij}) \rangle_{ca} = 
\int D(\psi^{\ast}\psi)\,\exp\left\{ A_{0}^{e\!f\!f} + 
A_{SG}^{replic}\right\}~~~~ 
\label{a5}
\end{eqnarray}
where

\begin{eqnarray} 
A_{0}^{e\!f\!f} =  \sum_{ij\sigma\alpha}\sum_{\omega_n}
\psi_{i\sigma\alpha}^{\ast}(\omega_n)
g_{ij}^{-1}(\omega_n)\psi_{j\sigma\alpha}(\omega_n) \nonumber
\end{eqnarray}
and

\begin{eqnarray} 
A_{SG}^{replic} = \frac{1}{N}\left\{  \frac{\beta^{2} J^{2}}{2}\sum_{\alpha\beta}
\left[\sum_{i}4S_{i}^{\alpha}S_{i}^{\beta}\right]^{2} +\,\, \right. \nonumber \\
  \left.\frac{\beta J_{0}}{2} \sum_{\alpha}\left[\sum_{i}2S_{i}^{\alpha}\right]^{2}\,\right\}.\nonumber
\end{eqnarray}
In the previous equation, \( g_{ij}^{-1}(\omega_n) \) has been defined in 
Section \ref{model}. The static approximation is also used to write the 
$ S_{i}^{\alpha}$ in terms of Grassmann fields as in equation (\ref{a4}) and the
resulting equation can be linearized by standard procedures \cite{r10}, 
introducing the order parameters \( q_{\alpha\beta}\) and \( m_{\alpha}\) 
which gives equations (\ref{e10}) and (\ref{e11}).


\end{document}